\documentclass[9pt,twocolumn,twoside]{opticajnl}
\journal{opticajournal} 

\setboolean{shortarticle}{false}

\usepackage{amsmath}
\usepackage{lineno}
\usepackage{layouts}
\usepackage{fancyhdr}
\usepackage{hyperref}
\usepackage{url}
\usepackage{siunitx}

\DeclareSIUnit{\samplepersecond}{S/s}

\title{Electro-optic time transfer with femtosecond stability}

\author[1,*]{Joshua Olson}
\author[2]{Robert Rockmore}
\author[3]{Nathan D. Lemke}
\author[2]{Sean Krzyzewski}
\author[2]{Brian Kasch}

\affil[1]{Space Dynamics Laboratory, Utah State University}
\affil[2]{Air Force Research Laboratory, Space Vehicles Directorate}
\affil[3]{Bethel University}
\affil[*]{joshua.l.olson@sdl.usu.edu} 

\begin{abstract} 
Optical two-way time and frequency transfer is an enabling technology that has applications ranging from fundamental investigations of relativity to the operation of global navigation satellite systems. While fiber frequency combs have demonstrated the most stable optical links, they are not ideal for applications that require very low SWaP-C. Here, we demonstrate two-way time and frequency transfer using electro-optic combs that have a direct path to full chip-scale integration. This two-way electro-optic time and frequency transfer system demonstrated instabilities as low as \SI{15}{\femto\second} at \SI{1}{\second} of averaging time. These results show a pathway to highly stable, agile and low SWaP-C time transfer networks.
\end{abstract}

\setboolean{displaycopyright}{false} 
\begin{document}
\maketitle
\section{Introduction}

\begin{figure*}[htbp]
    \centering\includegraphics[width=\textwidth]{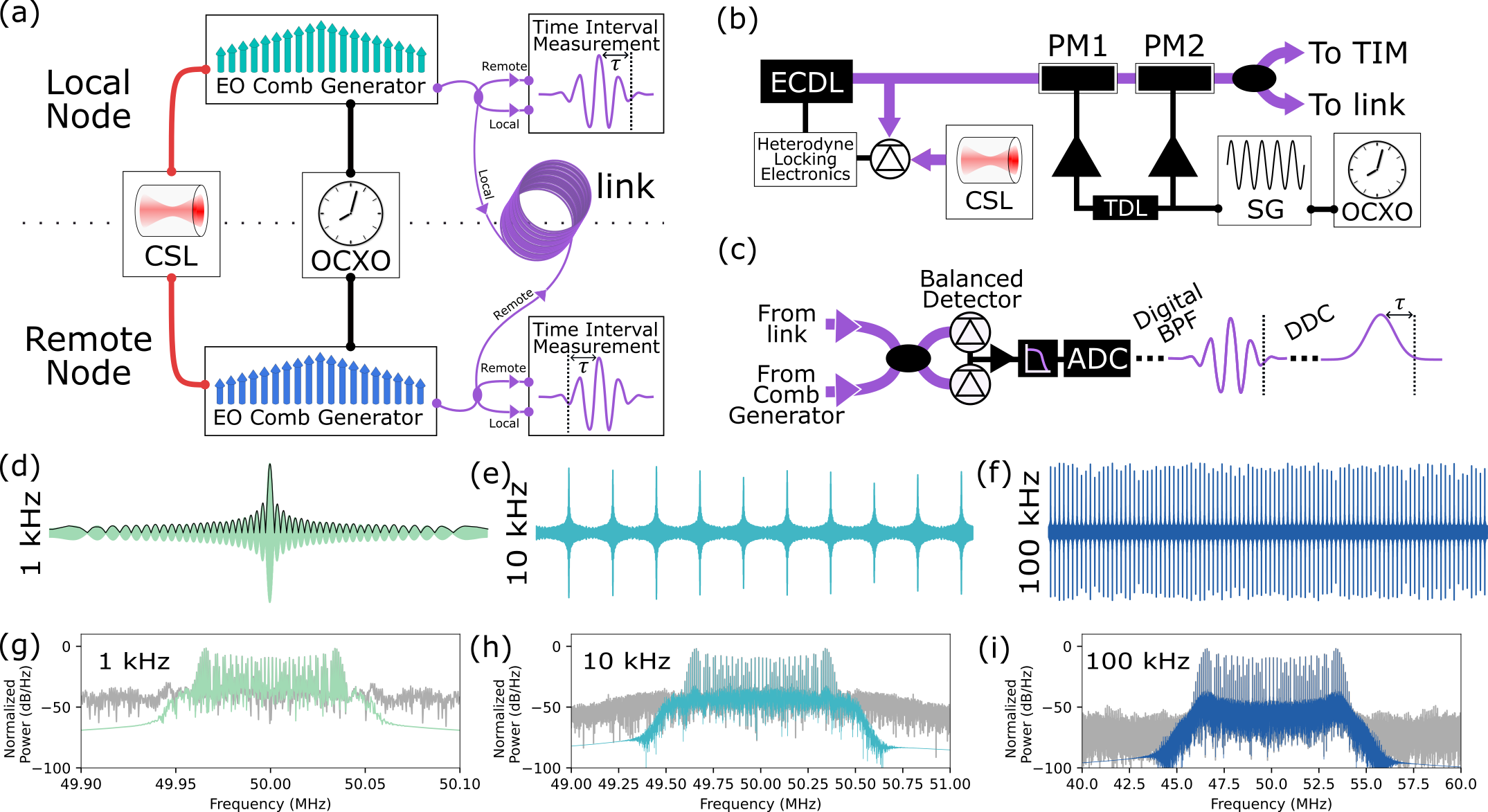}
    \caption{(a) Overview of the EOTT system. 
    Linear optical sampling between two EO combs is performed on two ends of a \SI{1}{\kilo\meter} fiber optical link.
    The EO combs are individually generated at the remote node and the local node and share both an optical reference from a common CSL and an RF reference from an OCXO.
    (b) Diagram of an EO comb generator. An external cavity diode-laser is locked to a CSL and then travels through two over-driven phase modulators (PM) to generate an EO comb.
    The phase modulators are driven by an amplified RF tone from an RF signal generator (SG).
    TDL, tuneable RF delay line; TIM, time interval measurement.
    (c) Diagram of the time-interval measurement system.
    Light from one of the EO comb generators is coupled together with light from the other EO comb generator that has traversed the optical link.
    The coupled light is directed to a balanced photodiode.
    After analog low-pass filtering the signal is digitized on an analog-to-digital converter.
    Digital band-pass filtering (BPD) increases the signal-to-noise ratio and digital-down conversion (DDC) removes the carrier frequency before peak detection is performed.
    (d-f) \SI{1}{\milli\second} time record of LOS interferograms produced with repetition rate differences of (d) \SI{1}{\kilo\hertz} (e) \SI{10}{\kilo\hertz} (f) \SI{100}{\kilo\hertz}. Time-records are shown after applying a digital band-pass filter.
    (g-i) Interferogram spectra with repetition rate differences of (g) \SI{1}{\kilo\hertz} (h) \SI{10}{\kilo\hertz} (f) \SI{100}{\kilo\hertz}.
    The interferogram spectra are shown both before digital band-pass filtering (gray) and after digital band-pass filtering (color).}\label{overview}
\end{figure*}

Comparison between distant optical clocks could lead to new understanding about the nature of gravity \cite{dereviankoFundamentalPhysicsStateoftheart2022,ste-questteamSTEQuestSpaceTimeExplorer2013}, improve global navigation satellite systems (GNSS) \cite{berceauSpacetimeReferenceOptical2016}, enable precise geodetic mapping \cite{grottiGeodesyMetrologyTransportable2018}, and reduce clock-based inaccuracies in very long baseline interferometry \cite{clivatiCommonclockVeryLong2020a, pizzocaroIntercontinentalComparisonOptical2021}.
Deployed optical clock networks would consist of highly stable optical references, along with highly precise optical frequency dividers and free-space optical time-transfer transceivers \cite{dereviankoFundamentalPhysicsStateoftheart2022,ste-questteamSTEQuestSpaceTimeExplorer2013,samainTimeTransferLaser2011}.
The significant cost of deploying such timing networks has also motivated the miniaturization and power reduction of key optical components.
With the demonstration of chip-scale components for optical references and stabilized optical clockwork \cite{newmanArchitecturePhotonicIntegration2019,moilleKerrInducedSynchronizationCavity2023} similarly high performing methods for comparing these optical clocks will be required that have a path to full photonic integration.

Two-way time and frequency transfer methods use timing signals travelling over reciprocal paths.
Departure and arrival times are measured independently at both local and remote sites and then compared to each other to remove the common time-delay of the reciprocal path.
This methodology has been well established using radio-frequency (RF) carrier signals to transfer time between satellites \cite{stoneTwoWaySatelliteTime2009}.
The accuracy of this method was improved by over an order of magnitude by using pulsed lasers as a timing signal \cite{samainTimeTransferLaser2011,schreiberGroundbasedDemonstrationEuropean2010}, reaching sub-picosecond stability over 100 s in a laboratory demonstration \cite{schreiberGroundbasedDemonstrationEuropean2010}.
Recent work has demonstrated sub-picoseond level stability using RF modulation of optical carrier signals.
Spread-spectrum modulation on an optical carrier was shown to achieve a stability of nearly \SI{200}{\femto\second} at \SI{1}{\second} and single picosecond performance beyond \SI{10}{\second} \cite{khaderTimeSynchronizationFreespace2018}. 
A comparable split-spectrum modulation technique was also demonstrated with a noise floor of \SI{1}{\pico\second} \cite{yangPicosecondprecisionOpticalTwoway2020}.
Time-transfer over an 86-km fiber link was performed with a noise floor of \SI{400}{\femto\second} using pulsed phase modulation of an optical carrier signal \cite{frankSubpsStabilityTime2018}.
To the best of our knowledge, the stability of RF over optical time-transfer methods are currently limited to a few hundreds of femtoseconds.

The most stable free-space optical two-way time and frequency transfer (O-TWTFT) links have been demonstrated using fully self-referenced optical frequency combs (OFC) locked to a stable optical frequency reference \cite{giorgettaOpticalTwowayTime2013,deschenesSynchronizationDistantOptical2016,bergeronTightRealtimeSynchronization2016,sinclairFemtosecondOpticalTwoway2019,shenExperimentalSimulationTime2021,shenFreespaceDisseminationTime2022,caldwellQuantumlimitedOpticalTime2023,martinDemonstrationRealTimePrecision2023}.
One O-TWTFT method uses linear optical sampling (LOS) between two frequency combs. In the LOS method, sub-femtosecond level timing information carried by the optical pulses is magnified in time and can be detected above the picosecond level photodetection jitter limit \cite{giorgettaOpticalTwowayTime2013,shenFreespaceDisseminationTime2022}.
A more photon-efficient comb based O-TWTFT method has recently been demonstrated, which uses real-time control of the repetition rate and carrier envelope offset frequency of fiber frequency combs to track the pulse arrival time from a distant fiber frequency comb \cite{caldwellQuantumlimitedOpticalTime2023}.
The fiber comb O-TWTFT methods have the potential of transferring timing information between very low noise optical atomic clocks \cite{dereviankoFundamentalPhysicsStateoftheart2022}.
However, the fiber-based comb sources in these highly stable time transfer systems also have fundamental limits to their size, weight, power and cost (SWaP-C).

OFCs can also be generated using direct EO modulation of an optical carrier with RF signals \cite{kobayashiHighrepetitionrateOpticalPulse1972,kourogiMonolithicOpticalFrequency1994,fujiwaraOpticalCarrierSupply2003,huangNonlinearlyBroadenedPhasemodulated2008,metcalfHighPowerBroadlyTunable2013,metcalfBroadlyTunableLow2015}.
These electro-optic (EO) combs are a compelling alternative to fiber-frequency combs in applications that can benefit from accessing a large range of repetition rates, fast frequency switching, predictable output over a wide temperature range, or a flat optical spectrum \cite{parriauxElectroopticFrequencyCombs2020b}.
As a result, EO comb generators have been implemented for gas sensing \cite{martin-mateosDualElectroopticOptical2015a,duranUltrafastElectroopticDualcomb2015a}, optical communications \cite{huChipbasedOpticalFrequency2021} and as frequency references for astronomical spectroscopy \cite{yiDemonstrationNearIRLinereferenced2016,kashiwagiDirectGeneration125GHzspaced2016}.
Although EO comb generators based on bulk modulators have similar SWaP-C to fiber-combs \cite{omalleyArchitectureIntegratedRF2023a}, they have a path to significant SWaP-C improvements through on-chip photonic integration \cite{wangIntegratedLithiumNiobate2018,heHighperformanceHybridSilicon2019a,renIntegratedLowVoltageBroadband2019,huHighefficiencyBroadbandOnchip2022,hanLowpowerAgileElectrooptic2024,zhangPowerefficientIntegratedLithium2023}.
These potential improvements and the ease with which the comb frequencies can be controlled make EO combs a compelling option for low SWaP-C optical time and frequency transfer.

In this work, we use LOS between two highly tunable bulk EO comb generators to demonstrate O-TWTFT over a \SI{1}{\kilo\meter} fiber optical link within a single laboratory.
This demonstration uses a single clock that is common to both the local and remote nodes of the optical link to isolate the additive timing noise of the EO comb based LOS process.
This technique employs RF modulation of an optical carrier, but we are able to achieve an improved noise floor over other RF over optical techniques by using LOS and low noise single tone RF synthesizers.
After cancellation of timing error induced by the optical link, stability below \SI{15}{\femto\second} is reached after \SI{1}{\second} of averaging.
\SI{15}{\femto\second} stability is maintained with link attenuation as high as \SI{30}{\dB}.
This proof-of-principle demonstration provides a compelling case for the use of two-way electro-optic time transfer (EOTT) for deployed time-transfer networks.

\section{Methods}

In this work, we demonstrate the two-way exchange of elapsed time-interval information between two nodes of our time and frequency transfer system separated by a \SI{1}{\kilo\meter} fiber-optical link. This time interval information is carried from the clock at each node by an electro-optic frequency comb generated from an RF modulated signal that is referenced to the node's clock.
The relative time interval information between the clocks are compared using LOS between the remote and local EO combs when the repetition rate of the local comb, $f_{r,L} = f_r$, and the repetition rate of the remote comb $f_{r,R} = f_r + \Delta f_r$ are offset by $\Delta f_r$.
Temporal walk-off between the two pulse trains produces a cross-correlation signal, a time-domain interferogram, that carries information about the relative timing and phase between the two frequency combs \cite{dorrerLinearOpticalSampling2003a, coddingtonCoherentLinearOptical2009}. This interferogram signal is detected at a rate equal to $\Delta f_r$, thus $\Delta f_r$ is the update rate of the time-transfer link.
Figure \ref{overview} (d-f) shows a record of the time-domain interferograms where $f_r$ was set to \SI{200}{\mega\hertz} and $\Delta f_r$ is tuned to \SI{1}{\kilo\hertz}, \SI{10}{\kilo\hertz} and \SI{100}{\kilo\hertz}.
The electrical field information of the combined combs is time magnified such that the down-converted field information can be measured with a \SI{100}{\mega\hertz} bandwidth photodiode and a \SI{200}{\mega\hertz} sample rate analog-to-digital converter (ADC). The amount of temporal magnification, M, is determined by the relative offset between the pulse repetition rates, $M = f_{r}/\Delta f_{r}$.

Measurement of the arrival time of the time-domain interferograms are made at both the local node, $\tau_L$, and at the remote node, $\tau_R$. These measurements include information about time-interval differences between the clocks at each node, $\tau_{LR}$, time-of-flight noise in the common-path optical link, $\tau_{TOF}$, and residual measurement errors, $\epsilon_{\tau,L/R}$:
\begin{equation}
\tau_L = \tau_{LR} - \tau_{TOF} + \epsilon_{\tau,L}
\end{equation}
\begin{equation}
    \tau_R = \tau_{LR} + \tau_{TOF} + \epsilon_{\tau,R}
\end{equation}
$\epsilon_{\tau,L/R}$ include both additive noise from the EO comb generators and technical noise from non-common optical paths at each node.
Assuming reciprocity of the link, $\tau_{TOF}$ is common to measurements of the interferogram arrival times made on both sides of the optical link.
The change in sign of $\tau_{TOF}$ is caused by the asymmetry between the repetition rates of the two combs.
Due to aliasing in the LOS down-conversion process a change in optical path length over the link will either advance or delay the arrival time of the interferograms depending on if the slow local comb, $f_{r,L} = f_r$, or the fast remote comb, $f_{r,R} = f_r + \Delta f_r$ traverses the optical link.
Consequently, the local node measurement of the link time-of-flight noise is anti-correlated to the remote node measurement of the link time-of-flight noise.
$\tau_{LR}$ is common to both measurements and positively correlated.
As a result, $\tau_{LR}$ can be isolated from $\tau_{TOF}$ by adding the time-interval measurements from the local and the remote node such that,
\begin{equation}\label{tickrate}
\frac{\tau_{L} + \tau_{R}}{2} = \tau_{LR} + \epsilon_{\tau}
\end{equation}
Here $\epsilon_{\tau}$ is the combined residual timing error from measurements at the local and remote sites and limits the total link stability.
Our system uses a common-clock configuration in which both the local and the remote nodes share the same RF reference, and therefore the clocks at the two nodes always share the same time-interval.
Therefore, $\tau_{LR} = 0$ and equation (\ref{tickrate}) will include noise from the electro-optic comb generators and non-common beam paths but will not include relative time interval differences between clocks at the local and remote nodes.
This provides a direct measurement of the additive noise of our electro-optic transfer comb apparatus.

The experimental apparatus is shown in Figure \ref{overview} (a).
The experiment includes two EO comb generators, shown in Figure \ref{overview} (b), one generator for the local node, and one for the remote node of our system.
The optical reference in this experiment is derived from a single-frequency laser diode locked to an ultra-stable optical cavity.
This cavity stabilized laser (CSL) is used as an optical reference to stabilize two additional low-noise diode lasers which are then used to generate the local and remote EO combs.
The RF reference is derived from a commercially available oven controlled crystal oscillator (OCXO).
This OCXO is used to synchronize (1) the offset locks between the comb generator diode lasers and the CSL (2) the RF signal that is used to generate sidebands around the EO comb carrier frequency and (3) the onboard clock of the digitizer used to measure the timing information from the two combs. Other than the CSL, the OCXO and the use of a single multichannel analog-to-digital converter, all the components in the two nodes of the system are completely independent.

Each EO comb is generated using an external cavity diode laser (ECDL) with a free-running linewidth that is less than 5 kHz.
The ECDLs are temperature tuned and are each mixed with the CSL on a separate balanced photodetector. Feedback from the beat signals are used to lock each ECDL by varying its laser current.
The remote comb ECDL is locked with a \SI{300}{\mega\hertz} offset from the CSL while the local comb ECDL is locked with a \SI{350}{\mega\hertz} offset from the CSL.

The optical frequency combs are generated by phase modulating the light from each ECDL in a series of bulk electro-optic (EO) phase modulators. Two EO phase modulators are used in each EO comb generator. Up to \SI{1}{\milli\watt} of average optical power is measured at the output of the generators. 
The phase modulators are driven by phase-lock loop (PLL) based RF synthesizers. The output from the RF synthesizers is split and amplified to \SI{1.6}{\watt}, measured before the input port of the EO modulators.
Figure \ref{PN_TDEV} (a) shows the measured phase noise of the RF synthesizers when generating a \SI{200}{\mega\hertz} tone.

The output from the local EO comb is split using fiber-optical couplers and part of the light is directed over a spooled \SI{1}{\kilo\meter} fiber optic cable (Figure \ref{overview} (a)).
At the other end of this fiber-optical link, at the remote-node, the light is combined with the remote EO comb signal using a 2x2 fiber-optical coupler with a 50 percent splitting ratio.
This combined local and remote light is then sent to another 50:50 fiber-optical coupler with both outputs from the coupler placed onto a balanced photodetector as shown in Figure \ref{overview} (c).
The light from the remote EO comb is also split and sent across the fiber-optical link in the opposite direction before being combined with the local EO comb and similarly detected on a balanced photodetector at the local node.

The RF signal measured on the photodiodes is low-pass filtered at \SI{100}{\mega\hertz}. Low pass-filtering is used to isolate the low-frequency interference signal, the time-domain interferogram, that contains information about the relative arrival times of the remote and local frequency combs on each side of the optical link. 
The interferograms are digitized on a 14-bit ADC with a sample-rate of \SI{200}{\mega\samplepersecond}.
For experimental convenience the two interferograms are digitized on two separate channels of the same digitizer.
The data from the digitizer is taken continuously and saved before processing.
Figure \ref{overview} (d-f) shows a record of the time-domain interferograms.
The time-record data is digitally band-pass filtered to remove out of band photodetector noise (Figure \ref{overview} (g-i)).
Approximately 100 comb lines remain after band-pass filtering.

\section{Results}
\begin{figure}[htbp]
    \centering\includegraphics[width=0.95\linewidth]{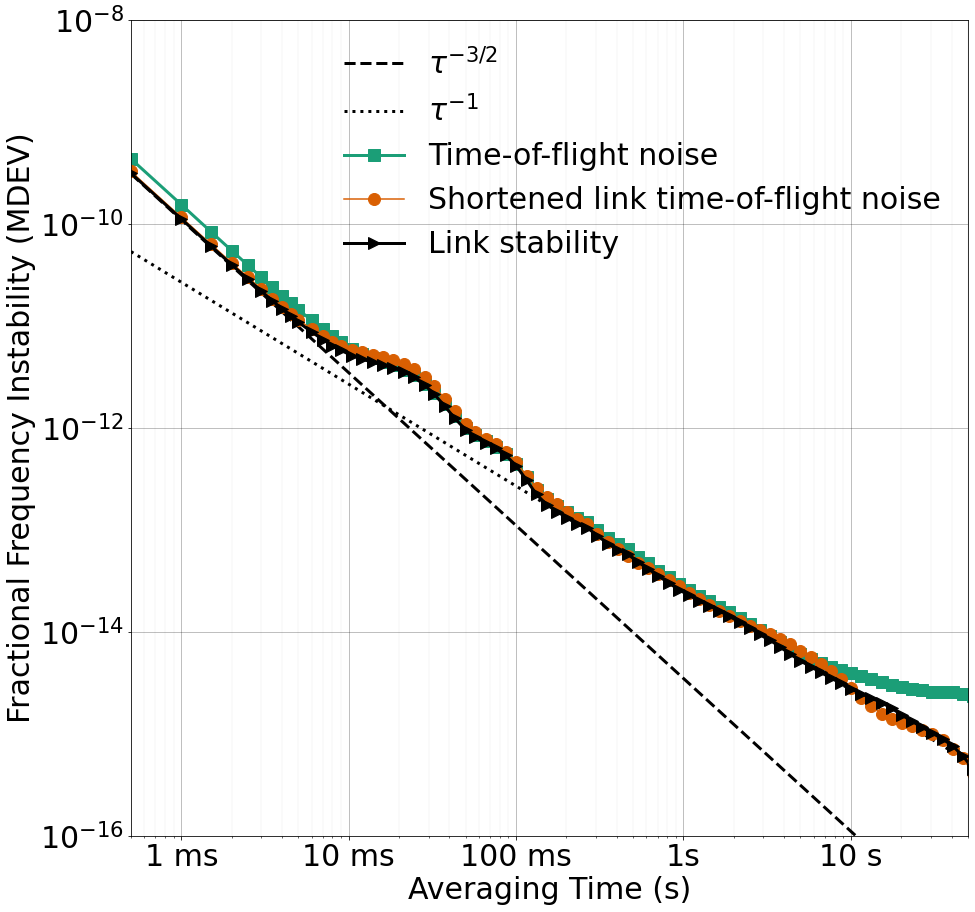}
    \caption{
    Modified Allan deviation (MDEV) measurement of the fractional frequency instability of the system.
    (Green, squares) Time-of-flight noise of the \SI{1}{\kilo\meter} fiber optic link shows the noise floor of a one-way time transfer system.
    (Orange, circles) Time-of-flight noise measurement over a shortened (1 m) fiber optic link shows that the increase in instability of the \SI{1}{\kilo\meter} time-of-flight above \SI{10}{\second} of averaging time is due to technical noise fluctuations in the long link.
    (Black, triangles) \SI{1}{\kilo\meter} link stability measurement using the two-way time-interval comparisons that removes common mode technical noise fluctuations of the long link.
    (Dashed) Expected MDEV slope for white phase noise.
    (Dotted) Expected MDEV slope for flicker phase noise.
    }
    \label{MDEV}
\end{figure}

\subsection*{Optical link stability}
Synchronizing two different clocks over an optical link generally requires time-interval measurements to be made in real-time and then shared across the optical link in order to make clock corrections or comparisons \cite{deschenesSynchronizationDistantOptical2016,martinDemonstrationRealTimePrecision2023}.
Due to the common clock configuration used in this system, clock synchronization is guaranteed, and real-time processing was not required.
Instead, gapless records of interferograms are saved to memory for each node of the system and the system noise floor is calculated during post-processing.

Figure \ref{MDEV} shows the measured two-way instability over the \SI{1}{\kilo\meter} optical link.
For this measurement, the local comb is generated with a \SI{200}{\mega\hertz} modulation frequency and the remote comb is generated with modulation frequency of \SI{200.002}{\mega\hertz}.
This produces a \SI{2}{\kilo\hertz} repetition rate difference ($\Delta f_r$) between the local and remote EO combs.
The two-way link reaches instability as low as $2.4\times10^{-14}$ at \SI{1}{\second}. At \SI{50}{\second}, the longest record length, instability as low as $5.5\times10^{-16}$ is measured.
Below \SI{5}{\milli\second} of averaging time the modified Allan deviation (MDEV) of the link decreases proportionally to $\tau^{-3/2}$ indicating that white phase noise is dominant over this time-scale.
Above \SI{5}{\milli\second} of averaging time the MDEV of the link decreases proportionally to $\tau^{-1}$, except for two peaks above \SI{25}{\milli\second}, which appears to be caused the PLLs in the RF signal generators. The $\tau^{-1}$ trend to the MDEV indicates that the system performance is limited by flicker phase noise above \SI{5}{\milli\second}.

Beyond \SI{7}{\second} of averaging time, the time-of-flight noise of the optical link dominates the one-way time-interval measurement between the local and the remote node.
The two-way measurement reduces this time-of-flight noise down to the $\tau^{-1}$ noise floor which also corresponds very closely to the same two-way measurement over a shortened \SI{1}{\meter} fiber-optical link.
The corresponding two-way time deviation measurement for the \SI{1}{\kilo\meter} link is shown in Figure \ref{PN_TDEV} (b), demonstrating a noise floor of just under \SI{15}{\femto\second} after \SI{1}{\second} of averaging.

\subsection*{Performance Limitations}

\begin{figure}[htbp]
    \centering\includegraphics[width=0.95\linewidth]{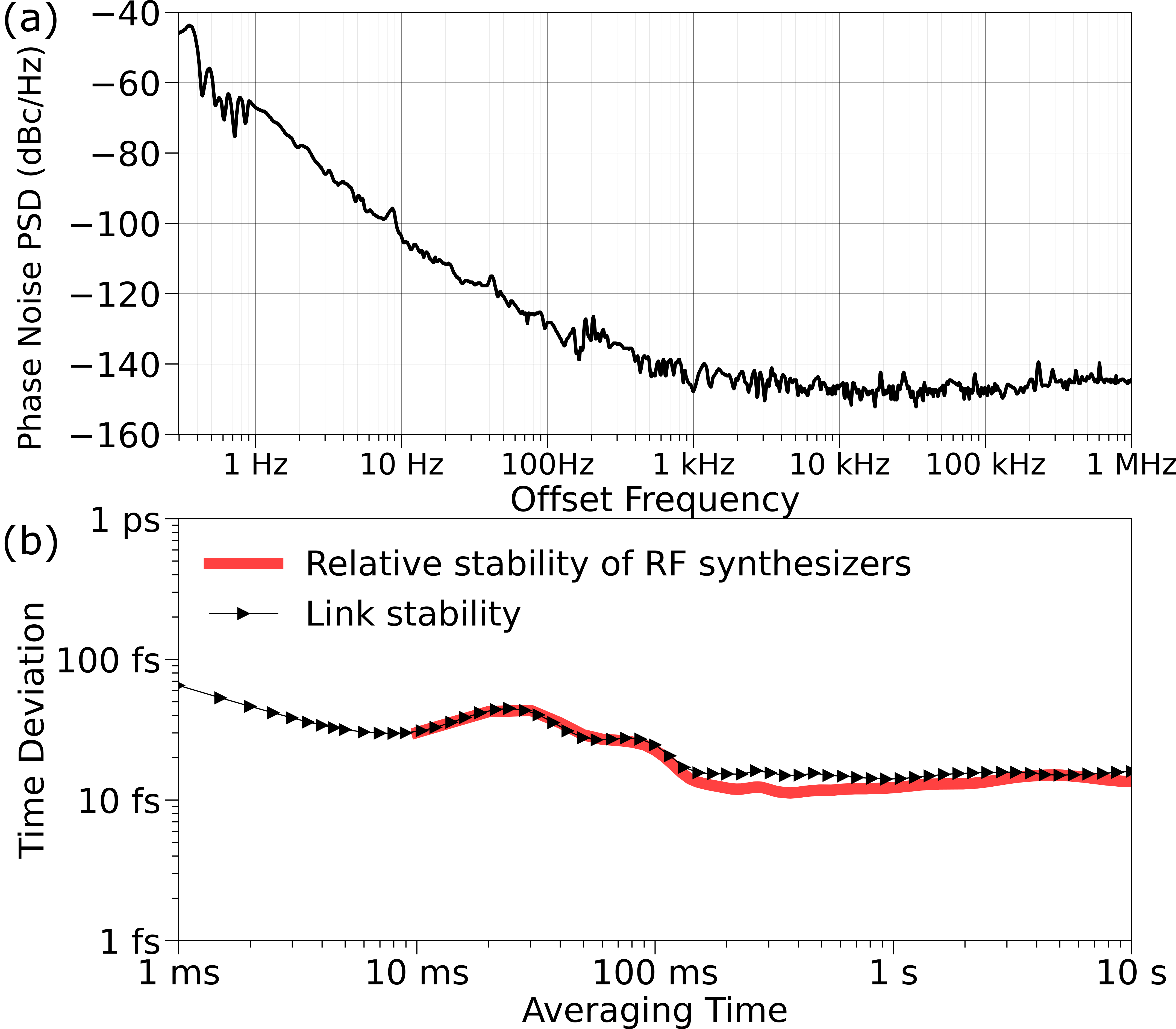}
    \caption{
    (a) Phase noise measurement of one of the RF synthesizers when generating a \SI{200}{\mega\hertz} tone.
    (b) Time deviation measurement of the link stability (black) compared to a measurement of the relative stability between the two RF synthesizers (red).
    The measured link stability is shown to closely match the relative stability between the RF synthesizers.
    }
    \label{PN_TDEV}
\end{figure}

The stability of the system at long averaging times is limited by the relative stability between the two RF synthesizers. The phase noise of a single synthesizer is shown in Figure \ref{PN_TDEV} (a). Measurement of the relative stability between the two synthesizers is made by mixing a \SI{200}{\mega\hertz} tone from one synthesizer with a \SI{210}{\mega\hertz} tone from the other synthesizer when both synthesizers shared a common reference signal. The phase of the resulting \SI{10}{\mega\hertz} difference frequency is measured using a phase noise analyzer. The raw phase measurements made at \SI{10}{\mega\hertz} are scaled to a \SI{2}{\kilo\hertz} carrier frequency for comparison to the measured link stability. Figure \ref{PN_TDEV} (b) shows the directly measured time deviation between the two synthesizers compared to the time deviation measured across the \SI{1}{\kilo\meter} link. The measurement of the link stability using LOS closely matches the relative stability of the synthesizers, including the peaks above \SI{25}{\milli\second}, indicating that the system performance is limited by flicker phase instabilities between the RF synthesizers above \SI{10}{\milli\second} of averaging time. Improvements in link stability may be possible by using optical methods of generating low-phase noise RF signals \cite{xiePhotonicMicrowaveSignals2017,kalubovilageXBandPhotonicMicrowaves2022}.

\begin{figure}[htbp]
    \centering\includegraphics[width=0.95\linewidth]{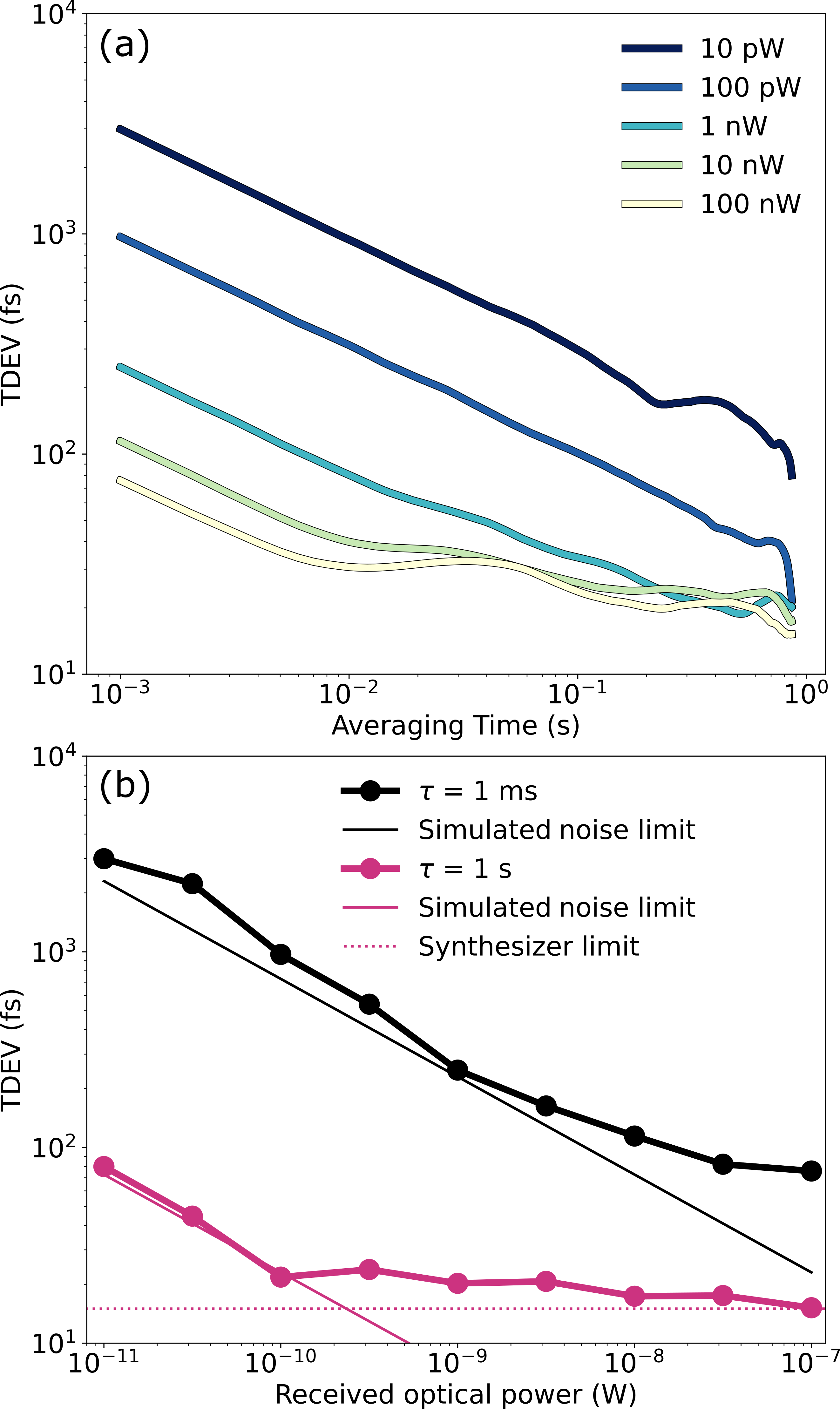}
    \caption{
        (a) Measurement of link time deviation while the link is operating at an update rate of \SI{1}{\kilo\hertz} while attenuating the link signal.
        Link time deviations are shown at received optical powers from \SI{10}{\pico\watt} to \SI{100}{\nano\watt} in \SI{10}{\dB} increments.
        (b) Link time deviation measured at different received optical powers.
        (black, dots) no averaging, (pink, dots) after \SI{1}{\second} of averaging.
        Photodetector noise limits are calculated from simulation (black, solid) no averaging, (pink, solid) 1s of averaging.
        (Pink, dotted) Noise floor from the relative stability of the two RF synthesizers at \SI{1}{\second} of averaging.
        }
    \label{attenuation}
\end{figure}

Over free-space optical links, the performance of the system would be subject to significant loss from propagating long distances through turbulent atmosphere \cite{jahidContemporarySurveyFree2022}.
To measure the performance of the system with high link loss, we added calibrated optical attenuators to the fiber optical link when operating with an update rate of \SI{1}{\kilo\hertz}.
We were able to add over 40 dB of attenuation to the link with transmitted signal powers as low as \SI{100}{\nano\watt} before degrading the performance of the system, as shown in Figure \ref{attenuation}(a).
With transmitted optical powers below \SI{100}{\nano\watt} the stability of the link begins to degrade due to a reduction of the signal-to-noise ratio of the measured interferograms.
Below 100 nW of received optical power the time deviation at \SI{1}{\milli\second} begins to increase, eventually increasing at a rate inversely proportional to the square root of the transmitted optical power: $\sigma\left(1~ms\right) \propto 1/\sqrt{P_{link}}$ (Figure \ref{attenuation}(a)).

To identify the noise source that caused this decreased stability, numerical simulation of the EO comb generators, optical link, and time-interval measurement was performed.
In experiment, noise from the amplified balanced photodetectors was measured to have a nearly flat spectrum with a standard deviation of \SI{1.9}{\milli\volt} after digitization.
The digitizer sampled this noise at a rate of \SI{200}{\mega\samplepersecond} and \SI{100}{\mega\hertz} low pass filters were used on the digitizer inputs.
In simulation, white amplitude noise with this standard deviation was added to simulated interferograms measured at the same received optical powers that were observed in the experiment.
Simulated interferogram arrival times were calculated using the same post-processing algorithm that was used to process the experimental data.
A linear fit to the simulated deviation data was used to determine the detector noise limit and is shown in figure \ref{attenuation}(b).
The simulated noise limit from photodetector noise closely matches the noise limited performance of the system under large link attenuation, indicating that photodetector noise dominates at low received optical powers.
Over long averaging times some of this noise can be averaged down to reach better performance.
Figure \ref{attenuation}(b) shows that after \SI{1}{\second} of averaging, the stability of the system with \SI{1}{\kilo\hertz} update rate is mostly unaffected with received optical powers above \SI{100}{\pico\watt}.
Below \SI{100}{\pico\watt} the link stability at \SI{1}{\second} also begins to follow the photodetector noise limit.
For very long distance links some of these attenuation issues could be offset by using optical amplifiers.
To test the performance of the system with optical amplification, an erbium doped fiber amplifier (EDFA) was added at the output of the EO comb generators.
The EDFA produced \SI{20}{\dB} of signal gain, corresponding to an output power of \SI{100}{\milli\watt}.
The system performance with the EDFA included showed no performance degradation from amplified spontaneous emission from the EDFA.
This indicates that optical amplification could be used to establish longer free space optical links potentially without decreasing the link stability.

\subsection*{Tunability of the repetition rate}
One advantage of using EO transfer combs is the ability to widely and quickly tune the comb modulation frequencies.
Adjusting the modulation frequency of a single comb produces a change in the repetition rate difference, $\Delta f_r$, between the two combs.
Tuning just $\Delta f_r$ produces a change in both the link update rate, $1/\Delta f_r$, the rate at which time-intervals can be measured via an increased rate of interferogram arrival, and a change in the dual-comb magnification factor.
Reduction of the magnification factor produces interferograms with shorter temporal duration and with a proportionally larger RF bandwidth.
Figure \ref{overview} (d-f) shows recorded interferograms at different values of $\Delta f_r$.

\begin{figure}[htbp]
    \centering\includegraphics[width=0.95\linewidth]{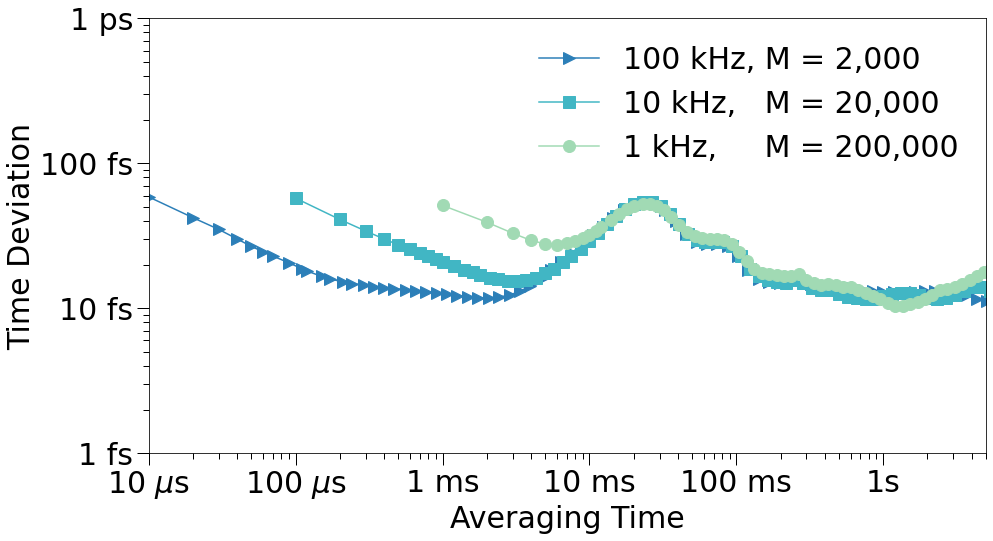}
    \caption{
    Measurement of the time-deviation over the \SI{1}{\kilo\meter} fiber optical link with different link updates rates (repetition rate differences, $\Delta f_r$).
    (Triangles) \SI{100}{\kilo\hertz} updates rate (Squares) \SI{10}{\kilo\hertz} update rate (Circles) \SI{1}{\kilo\hertz} update rate.
    }
    \label{TDEV_tuning}
\end{figure}

Time intervals were measured at both the local site and the remote site when the local comb was set to a repetition rate of \SI{200}{\mega\hertz} and the remote comb's repetition rate is tuned to \SI{200.001}{\mega\hertz}, \SI{200.010}{\mega\hertz} and \SI{200.100}{\mega\hertz} corresponding to repetition rate differences of \SI{1}{\kilo\hertz}, \SI{10}{\kilo\hertz} and \SI{100}{\kilo\hertz} respectively (Figure \ref{TDEV_tuning}).
Adjustments to $\Delta f_r$ do not significantly impact the measured time deviation at the fastest link update time, $\tau = 1/\Delta f_r$.
As a result, increasing $\Delta f_r$ leads to lower measured time deviations when white noise is dominant due to the ability to average down the white noise with more samples.
With a time interval update rate of \SI{100}{\kilo\hertz} the measured time deviation reaches \SI{15}{\femto\second} with only a few milliseconds of averaging.
Above \SI{1}{\milli\second} there is no observed performance benefit from higher link update rates as the instabilities quickly average down to the flicker phase noise floor from the signal generators.

\section{Conclusion}
In this work, we demonstrate a system for making two-way time interval measurements between clocks using LOS between electro-optic frequency combs.
This EOTT method demonstrated a link stability of \SI{15}{\femto\second} with \SI{1}{\second} of averaging in a common clock configuration.
Although this method cannot currently reach the stability of fiber-comb time-transfer methods, sub-fs clock synchronization may not be required for many applications.
Deployed time-transfer transceivers will be extremely sensitive to size, weight, power and cost.
For these applications EOTT, with components that can be fully integrated on chip, has significant benefits.
Furthermore, the wide frequency tuning range and frequency agility of EO comb generators can also provide a more flexible architecture for future time-transfer networks.

\begin{backmatter}
\bmsection{Funding}
Air Force Research Laboratory

\bmsection{Acknowledgments}
We thank Myles Silfies, Kyle Martin, Nader Zaki and Steve Lipson for their careful reading of this manuscript.

\bmsection{Disclosures}
The authors declare no conflict of interest. 

\bmsection{Disclaimers}
The views expressed are those of the author and do not necessarily reflect the official policy or position of the Department of the Air Force, the Department of Defense, or the U.S. Government.

\bmsection{Data Availability Statement}
Data underlying the results in this paper are not publically available at this time but may be obtained from the authors upon reasonable request.

\end{backmatter}

\bibliography{EOTT}

\end{document}